\documentclass[prb,aps,twocolumn]{revtex4}

\usepackage{amsmath}
\usepackage{bm}
\usepackage{graphicx}

\begin{document}

\title{Kibble--Zurek mechanism in a trapped ferromagnetic Bose--Einstein
condensate}

\author{Hiroki Saito$^1$, Yuki Kawaguchi$^2$, and Masahito Ueda$^3$}

\address{$^1$Department of Engineering Science, University of
Electro-Communications, Tokyo 182-8585, Japan \\
$^2$Department of Applied Physics and Quantum-Phase Electronics Center,
University of Tokyo, Tokyo 113-0032, Japan \\
$^3$Department of Physics, University of Tokyo, Tokyo 113-0033, Japan
}

\begin{abstract}
Spontaneous spin vortex formation in the magnetic phase transition of a
trapped spin-1 Bose--Einstein condensate is investigated using mean-field
theory.
In a harmonic trapping potential, an inhomogeneous atomic density leads to
spatial variations of the critical point, magnetization time scale, and
spin correlation length.
The Kibble--Zurek phenomena are shown to emerge even in such inhomogeneous
systems, when the quench of the quadratic Zeeman energy is fast enough.
For slow quench, the magnetized region gradually expands from the center
of the trap pushing out spin vortices, which hinders the Kibble--Zurek
mechanism from occurring.
A harmonic trap with a plug potential is also taken into account.
\end{abstract}

\maketitle

\section{Introduction}

Symmetry breaking phase transitions are considered to play crucial roles
in the early universe.
As the hot universe cooled down, the phase transitions broke the
symmetries of the vacuum fields.
Since causally disconnected regions acquire independent values of the
order parameter in the course of the phase transition, topological defects
can be left behind~\cite{Kibble}, such as monopoles, strings, and domain
walls.
It was proposed that this cosmological scenario of topological defect
formation can be tested by the normal fluid--superfluid phase transition
of liquid helium~\cite{Zurek}.
Such a mechanism of topological defect formation is called Kibble--Zurek
(KZ) mechanism, which has been studied in a wide variety of
systems~\cite{Chuang,Bowick,Hendry,Ruutu,Bauerle,Ducci,Carmi,Monaco,Maniv}.

Bose--Einstein condensates (BECs) of atomic gases are highly controllable
quantum systems and suitable for studying the KZ mechanism in a controlled
manner.
The BEC transition breaks the U(1) symmetry for a single-component system,
and quantized vortices can be formed by the KZ mechanism.
This has been demonstrated in the experiments reported in
Refs.~\cite{Scherer,Weiler}.
Spinor BECs (BECs of atoms with spin degrees of freedom) have a rich
variety of magnetic phases with different symmetry groups, and thus have
various kinds of topological defects~\cite{Kawaguchi,Kurn}.
In the experiments reported in Refs.~\cite{Sadler,Leslie}, the transition
from the polar state to the ferromagnetic state in a spin-1
$^{87}{\rm Rb}$ was observed, which was controlled by an external magnetic
field.
Formation of spin vortices by the KZ mechanism in this magnetic transition
has been investigated in
Refs.~\cite{Lamacraft,Uhlmann,Damski,SaitoKZ}.
It is predicted that the KZ mechanism can also be tested by the Mott
transition of cold atoms in an optical lattice~\cite{Dziarmaga}, soliton
formation in the BEC transition in a one-dimensional gas~\cite{Witkowska}, 
a miscible--immiscible transition in a binary BEC~\cite{Sabbatini}, and a
magnetic transition in an antiferromagnetic spinor BEC~\cite{Swislocki}.

In Ref.~\cite{SaitoKZ}, we studied the KZ mechanism in the magnetic
transition of a spin-1 $^{87}{\rm Rb}$ BEC and numerically demonstrated
the KZ scaling properties.
However, the numerical simulations in Ref.~\cite{SaitoKZ} were restricted
to the systems with uniform atomic density.
In the present paper, we perform numerical simulations of the
magnetization dynamics of a spin-1 BEC confined in a harmonic trapping
potential to show that the KZ mechanism can be observed in realistic
experiments.
We will show that the inhomogeneity of the trapped system has two effects
on the KZ properties.
The first one is caused by the spatial dependence of the spin correlation
length.
The number of spin vortices created by the KZ mechanism depends on the
spin correlation length, and therefore, depends on the position.
The second one originates from the competition between two velocities.
Since the density is high around the center of the atomic cloud, the
magnetization starts from the center and the magnetized region expands
outward.
If this expansion velocity is slower than the velocity of the spin wave,
the magnetized region can be causally connected with the region that is
going to magnetize, and the KZ mechanism breaks down.
A plug potential applied to the center of the trap is shown to resolve
this problem.

This paper is organized as follows.
Section~\ref{s:bogo} formulates the problem and provides mean-field and
Bogoliubov analyses.
Sections~\ref{s:sudden} and \ref{s:slow} show the numerical results for 
sudden quench and gradual quench of the magnetic field, respectively.
Section~\ref{s:plug} examines the case of a harmonic potential with a plug
potential.
Section~\ref{s:conc} concludes this paper.

\section{Mean-field analysis of spin correlations}
\label{s:bogo}

We consider bosonic atoms with mass $M$ and hyperfine spin $F = 1$
confined in an external potential $V_{\rm trap}(\bm{r})$.
The magnetic field $B$ is applied in the $z$ direction, and the linear and
quadratic Zeeman effects change the energies of spin sublevels $m = \pm 1$
by
\begin{equation}
p = \mp g_F \mu_{\rm B} B, \qquad
q = \frac{\mu_{\rm B}^2 B^2}{4 E_{\rm hf}},
\end{equation}
respectively, where $g_F$ is the hyperfine $g$ factor, $\mu_{\rm B}$ is
the Bohr magneton, and $E_{\rm hf}$ is the hyperfine splitting energy.
For $^{87}{\rm Rb}$ atoms, $g_F = 1/2$ for $F = 1$ and $E_{\rm hf} / h
\simeq 6.8$ GHz.
The interaction between atoms is characterized by spin-independent and
spin-dependent interaction coefficients given by
\begin{equation}
c_0 = \frac{4 \pi \hbar^2}{M} \frac{a_0 + 2 a_2}{3}, \qquad
c_1 = \frac{4 \pi \hbar^2}{M} \frac{a_2 - a_0}{3},
\end{equation}
respectively, where $a_S$ is the $s$-wave scattering length for two
colliding atoms with total spin $S$.
We use the values of $a_0 = 101.8 a_{\rm B}$ and $a_2 = 100.4 a_{\rm
B}$~\cite{Kempen} for $F = 1$ $^{87}{\rm Rb}$ atoms, where $a_{\rm B}$ is
the Bohr radius.

We employ the mean-field theory at zero temperature.
The state of the system is described by the macroscopic wave functions
$\psi_m(\bm{r}, t)$.
The mean-field energy is given by
\begin{eqnarray} \label{E}
E & = & \int d\bm{r} \biggl[ \sum_m \psi_m^* \left( -\frac{\hbar^2}{2M}
\nabla^2 + V_{\rm trap} + m p + m^2 q \right) \psi_m
\nonumber \\
& & + \frac{c_0}{2} \rho^2 + \frac{c_1}{2} \bm{F} \cdot \bm{F} \biggr],
\end{eqnarray}
where
\begin{eqnarray}
\rho(\bm{r}, t) & = & |\psi_1|^2 + |\psi_0|^2 + |\psi_{-1}|^2, \\
\bm{F}(\bm{r}, t) & = & \sum_{m,m'} \psi_m^* \bm{f}_{mm'} \psi_{m'},
\end{eqnarray}
with $\bm{f} = (f_x, f_y, f_z)$ being the spin-1 matrices.
The dynamics is given by the Gross-Pitaevskii (GP) equation,
\begin{equation} \label{GP}
i \hbar \frac{\partial \psi_m}{\partial t} = \frac{\delta E}{\delta
\psi_m^*},
\end{equation}
where the right-hand side indicates functional derivative.
In the rotating frame of the spin space ($\psi_{\pm 1} \rightarrow
e^{\mp i p t / \hbar} \psi_{\pm 1}$), the linear Zeeman terms in
Eq.~(\ref{GP}) can be eliminated, and we neglect them in the following
calculations.

To study the behaviors of the system analytically, we consider a uniform
system with density $\rho = n_0$.
When $c_1 < 0$ and $q > 0$, which is the case of spin-1 $^{87}{\rm Rb}$,
the ground state of Eq.~(\ref{E}) satisfying $F_z = 0$ is given by
\begin{equation} \label{polar}
\left( \begin{array}{c} \psi_1 \\ \psi_0 \\ \psi_{-1} \end{array} \right)
= \sqrt{n_0} e^{i \alpha} \left( \begin{array}{c} 0 \\ 1 \\ 0 \end{array}
\right)
\end{equation}
for $q > q_{\rm c}$ and
\begin{equation} \label{ba}
\left( \begin{array}{c} \psi_1 \\ \psi_0 \\ \psi_{-1} \end{array} \right)
= \sqrt{n_0} e^{i \alpha} \left( \begin{array}{c}
e^{i \beta} \frac{1}{2} \sqrt{1 - \frac{q}{q_{\rm c}}} \\
\frac{1}{\sqrt{2}} \sqrt{1 + \frac{q}{q_{\rm c}}} \\
e^{-i \beta} \frac{1}{2} \sqrt{1 - \frac{q}{q_{\rm c}}} \end{array} \right)
\end{equation}
for $q \leq q_{\rm c}$, where
\begin{equation}
q_{\rm c} = 2 |c_1| n_0,
\end{equation}
and $\alpha$ and $\beta$ are arbitrary phases.
The states in Eqs.~(\ref{polar}) and (\ref{ba}) are called the polar state
and broken axisymmetry state~\cite{Murata}, respectively.
The transverse magnetization of the polar state (\ref{polar}) is $(F_x^2 +
F_y^2)^{1/2} = 0$ and that of the broken axisymmetry state (\ref{ba})
is $(F_x^2 + F_y^2)^{1/2} = (1 - q^2 / q_{\rm c}^2)^{1/2}$.

We study the stability of the polar state (\ref{polar}) using the
Bogoliubov analysis.
Substituting
\begin{eqnarray}
\psi_0(\bm{r}, t) & = & e^{-i c_0 n_0 t / \hbar} \sqrt{n_0}, \\
\label{psi1}
\psi_{\pm 1}(\bm{r}, t) & = & e^{-i c_0 n_0 t / \hbar} \sum_{\bm k}
\frac{1}{\sqrt{V}} e^{i\bm{k} \cdot \bm{r}} a_{\pm 1, \bm{k}}(t),
\end{eqnarray}
into the GP equation (\ref{GP}), where $V$ is the volume of the system,
and keeping the first order terms in $a_{\pm 1, \bm{k}}$, we obtain
\begin{equation}
i \hbar \frac{d a_{\pm 1, \bm{k}}(t)}{dt} = (\varepsilon_k + q + c_1 n_0)
a_{\pm 1, \bm{k}}(t) + c_1 n_0 a_{\mp 1, -\bm{k}}^*(t),
\end{equation}
where $\varepsilon_k = \hbar^2 k^2 / (2M)$.
The solution is given by
\begin{eqnarray} \label{sol}
a_{\pm 1, \bm{k}}(t) & = & \left( \cos \frac{E_k t}{\hbar} - i
\frac{\varepsilon_k + q + c_1 n_0}{E_k} \sin \frac{E_k t}{\hbar} \right)
a_{\pm 1, \bm{k}}(0)
\nonumber \\
& & - \left( i \frac{c_1 n_0}{E_k} \sin \frac{E_k t}{\hbar} \right)
a_{\mp 1, -\bm{k}}^*(0),
\end{eqnarray}
where
\begin{equation} \label{ek}
E_k = \sqrt{(\varepsilon_k + q)(\varepsilon_k + q - q_{\rm c})}.
\end{equation}
When $q \geq q_{\rm c}$, $E_k$ is real for all $\bm{k}$, and
Eq.~(\ref{sol}) is an oscillating function.
In this case, the polar state (\ref{polar}) is stable against small
deviations.
When $q < q_{\rm c}$, $E_k$ is imaginary for $0 < \varepsilon_k <
q_{\rm c} - q$.
The modes with imaginary $E_k$ exponentially grow, which make the polar
state (\ref{polar}) dynamically unstable.

If the initial state is prepared in the stable polar state (\ref{polar})
with $q \geq q_{\rm c}$ and $q$ is decreased to $q < q_{\rm c}$, the
system becomes dynamically unstable and the transverse magnetization
$F_\pm$ emerges~\cite{Sadler,Leslie}.
Using Eq.~(\ref{psi1}) with Eq.~(\ref{sol}), the correlation function of
the transverse magnetization $F_\pm = F_x \pm i F_y$ is calculated to be
\begin{eqnarray} \label{FF}
& & \langle F_+(\bm{r}, t) F_-(\bm{r}', t) \rangle
\nonumber \\
& & = \frac{2n_0}{V}
\sum_{\bm{k}} \left| \cos \frac{E_k t}{\hbar} + i
\frac{\varepsilon_k + q}{E_k} \sin \frac{E_k t}{\hbar} \right|^2
e^{-i \bm{k} \cdot (\bm{r} - \bm{r}')}
\nonumber \\
& & \times \left[ \langle |a_{1, \bm{k}}(0)|^2 \rangle +
\langle |a_{-1, -\bm{k}}(0)|^2 \rangle \right],
\end{eqnarray}
where $\langle \cdots \rangle$ indicates the average with respect to
different initial values $a_{\pm 1, \pm\bm{k}}(0)$.
They include quantum and thermal fluctuations, residual atoms in the $m =
\pm 1$ states, and other experimental noises, and therefore we assume that 
$a_{\pm 1, \pm\bm{k}}(0)$ are independent complex random numbers.
For the dynamically unstable modes, Eq.~(\ref{FF}) contains the
exponentially growing factor $\exp(2 |E_k| t / \hbar)$, which has a sharp
peak at the most unstable wave number $k_{\rm mu}$.
We can thus approximate Eq.~(\ref{FF}) as~\cite{SaitoKZ}
\begin{eqnarray} \label{FF2}
& & \langle F_+(\bm{r}, t) F_-(\bm{r}', t) \rangle 
\nonumber \\
& & \propto \int d{\bm k}
\exp\left[ \frac{t}{\tau} \left( 1 - \frac{1}{4} \xi_{\rm corr}^2
\Delta k^2 \right) + i \bm{k} \cdot (\bm{r} - \bm{r}') \right],
\nonumber \\
\end{eqnarray}
where $\Delta k = k - k_{\rm mu}$, and $\tau$ and $\xi_{\rm corr}$ are
defined by
\begin{equation}
\frac{2|E_k|t}{\hbar} = \frac{t}{\tau} \left( 1 - \frac{1}{4}
\xi_{\rm corr}^2 \Delta k^2 \right) + O(\Delta k^4).
\end{equation}

For $q_{\rm c} / 2 < q < q_{\rm c}$, the most unstable wave number is
$k_{\rm mu} = 0$ with
\begin{eqnarray}
\tau & = & \frac{\hbar}{2 \sqrt{q(q_{\rm c} - q)}}, \label{tau1} \\
\xi_{\rm corr} & = & \sqrt{\frac{\hbar^2}{M} 
\frac{2q - q_{\rm c}}{q(q_{\rm c} - q)}}. \label{xi1}
\end{eqnarray}
In this case, the integral in Eq.~(\ref{FF2}) for a two-dimensional system
can be performed to yield
\begin{equation}
\langle F_+(\bm{r}, t) F_-(\bm{r}', t) \rangle \propto
\exp\left( \frac{t}{\tau} - \frac{\tau |\bm{r} - \bm{r}'|^2}
{t \xi_{\rm corr}^2} \right).
\end{equation}
For $q < q_{\rm c} / 2$, the most unstable wave number is
\begin{equation}
k_{\rm mu} = \sqrt{\frac{2M}{\hbar^2} \left( \frac{q_{\rm c}}{2} - q
\right)}
\end{equation}
with
\begin{eqnarray}
\tau & = & \frac{\hbar}{q_{\rm c}}, \label{tau2} \\
\xi_{\rm corr} & = & \sqrt{\frac{8\hbar^2}{M}
\frac{q_{\rm c} - 2q}{q_{\rm c}^2}}. \label{xcorr2}
\end{eqnarray}
The integral in Eq.~(\ref{FF2}) for a two-dimensional system becomes
\begin{equation}
\langle F_+(\bm{r}, t) F_-(\bm{r}', t) \rangle \propto \int k J_0(k
|\bm{r} - \bm{r}'|) e^{-\frac{t}{4\tau} \xi_{\rm corr}^2 \Delta k^2} dk,
\end{equation}
where $J_0$ is the Bessel function.
For $k_{\rm mu} |\bm{r} - \bm{r}'| \gg 1$, this expression can be
evaluated to be
\begin{eqnarray} \label{FF0}
& & \langle F_+(\bm{r}, t) F_-(\bm{r}', t) \rangle
\nonumber \\
& & \sim \cos(k_{\rm mu} r - \pi / 4) \exp\left( \frac{t}{\tau}
- \frac{\tau |\bm{r} - \bm{r}'|^2} {t \xi_{\rm corr}^2} \right).
\end{eqnarray}

\section{Numerical results}
\label{s:num}

We restrict ourselves to two-dimensional (2D) systems confined in a
harmonic potential $V_{\rm trap} = M \omega^2 (x^2 + y^2) / 2$ with
$\omega / (2\pi) = 2$ Hz.
When the system is tightly confined in the $z$ direction and the thickness
of the cloud is smaller than the spin healing length (typically a few
micrometers), the spin dynamics are effectively 2D.
We assume that the thickness in the $z$ direction is $\simeq 1$
$\mu{\rm m}$, and use 2D interaction coefficients as $c_j^{\rm 2D} = c_j
/ (1 \mu{\rm m})$.

We numerically solve the 2D GP equation using the pseudospectral
method~\cite{Recipes}.
The initial state is the ground state of Eq.~(\ref{E}) with $\psi_{\pm 1}
= 0$, which is obtained by the imaginary time propagation method.
We add small random noises to the initial state of $\psi_{\pm 1}$ to
trigger the magnetization.
We take a sufficiently large space so that the boundary condition does not
affect the results.

We define the transverse and longitudinal autocorrelation functions as
\begin{equation}
G_{\rm T} = \frac{\int |F_+|^2 d\bm{r}}{\int \rho^2 d\bm{r}}, \qquad
G_{\rm L} = \frac{\int F_z^2 d\bm{r}}{\int \rho^2 d\bm{r}}.
\end{equation}
We also define the transverse autocorrelation function along a circle with
radius $r$ as
\begin{equation} \label{gtr}
G_{\rm T}(r) = \frac{\int_0^{2\pi} |F_+|^2(r, \theta) d\theta}
{\int_0^{2\pi} \rho^2(r, \theta) d\theta}.
\end{equation}
The transverse spin winding number along a circle with radius $r$ is
defined as
\begin{equation} \label{wr}
w(r) = \frac{1}{2\pi} \int_0^{2\pi}
\frac{\partial}{\partial \theta}  {\rm arg} F_+(r, \theta) d\theta.
\end{equation}

\subsection{Sudden quench}
\label{s:sudden}

\begin{figure}[tb]
\includegraphics[width=7.5cm]{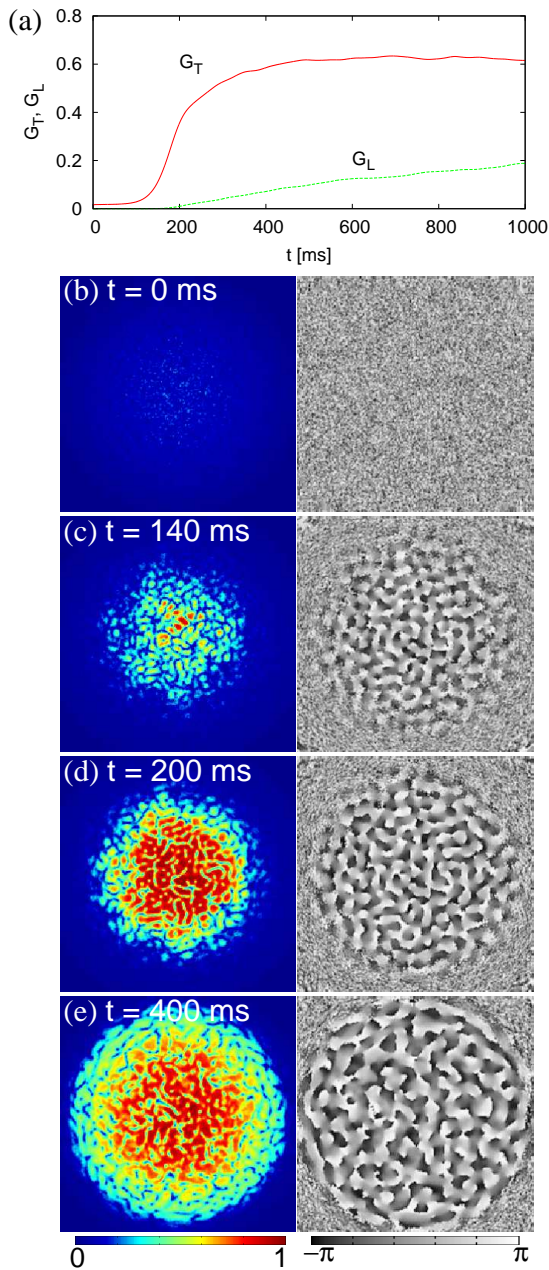}
\caption{
(a) Time evolutions of the autocorrelation functions $G_{\rm T}(t)$ and
$G_{\rm L}(t)$ for sudden quench to $q = 0$.
(b)--(e) Snapshots of the transverse magnetization $|F_+(\bm{r}, t)|$
(left panels) and its direction ${\rm arg} F_+(\bm{r}, t)$ (right
panels).
The unit of $|F_+(\bm{r}, t)|$ is $1.2 \times 10^{14}$ ${\rm m}^{-2}$.
The field of view of each panel is $400 \times 400$ $\mu {\rm m}$.
The number of atoms is $N = 10^7$.
See the supplementary data file for the movie showing the dynamics.
}
\label{f:sudden}
\end{figure}
We first investigate the magnetization dynamics for sudden quench of the
quadratic Zeeman energy to $q = 0$.
This corresponds to the situation in which the stable polar state is
prepared at sufficiently large $q$, and the magnetic field is suddenly
switched off at $t = 0$.
Figure~\ref{f:sudden}(a) shows the time evolution of the autocorrelation
functions $G_{\rm T}(t)$ and $G_{\rm L}(t)$.
The transverse magnetization starts to grow at $t \simeq 100$ ms and the
longitudinal magnetization follows.
Figures~\ref{f:sudden}(b)--\ref{f:sudden}(e) show the profiles of the
transverse magnetization.
The transverse magnetization emerges around the center and grows outward.
This is because the growth time in Eq.~(\ref{tau2}) is inversely
proportional to the atomic density and the magnetization grows fast at
which the density is large.
The total density distribution $\rho(\bm{r})$ is almost unchanged during
the time evolution, since $c_0$ is much larger than $c_1$.

Many spin vortices can be seen in
Figs.~\ref{f:sudden}(c)--\ref{f:sudden}(e) (the holes in the $|F_+|$
profiles, around which ${\rm arg} F_+$ rotate by $\pm 2\pi$).
In terms of the spin components in Eq.~(\ref{ba}), $\beta$ changes by $\pm
2\pi$ around the vortex core, which is occupied by the $m = 0$ component.
Such a spin vortex is called a polar-core vortex.
The spin winding number $w(r)$ defined in Eq.~(\ref{wr}) represents the
difference between the numbers of polar-core vortices with opposite
circulations within the radius $r$.

We note that the spin vortices are produced by two distinct mechanisms in
Fig.~\ref{f:sudden} with $q = 0$:
the KZ mechanism and the spin conservation
dynamics~\cite{SaitoKZ,Saito07}.
Since the spin correlation function in Eq.~(\ref{FF0}) has a finite
correlation length $\xi_{\rm corr}$, the directions of magnetization at
$\bm{r}$ and $\bm{r}'$ are independent for $|\bm{r} - \bm{r}'| \gg
\xi_{\rm corr}$, giving rise to the KZ mechanism.
On the other hand, when $q = 0$, the total magnetization $\int \bm{F}
d\bm{r}$ must be conserved at zero, since Eq.~(\ref{E}) is invariant with
respect to spin rotation in the rotating frame $\psi_{\pm 1} \rightarrow
e^{\mp i p t / \hbar} \psi_{\pm 1}$.
In the present case, however, the conservation law is more strict because
of the finite spin correlation length.
Since the spin directions at $\bm{r}$ and $\bm{r}'$ are independent for
$|\bm{r} - \bm{r}'| \gg \xi_{\rm corr}$, not only the total magnetization
but also the local magnetization $\int_{\rm local} \bm{F} d\bm{r}$
integrated over the size of $\sim \xi_{\rm corr}$ must be conserved in
each spatial region.
The magnetization thus occurs in such a way that the local magnetization
is conserved at zero, i.e., spin textures are formed~\cite{Saito05}.
Among various spin textures, the polar-core vortices are favorable, since
the excess energy at the defect can be minimized~\cite{Saito06}.
This is the second mechanism of the spin vortex formation in
Fig.~\ref{f:sudden}.
Thus, to see the effect of the KZ mechanism, we must take spatial region
much larger than $\xi_{\rm corr}$.
The correlation length is $\xi_{\rm corr} \simeq 10$ $\mu {\rm m}$ around
the center of the trap and $\xi_{\rm corr} \simeq 20$ $\mu {\rm m}$ at $r
= 200$ $\mu {\rm m}$ for Fig.~\ref{f:sudden}.

\begin{figure}[tb]
\includegraphics[width=8cm]{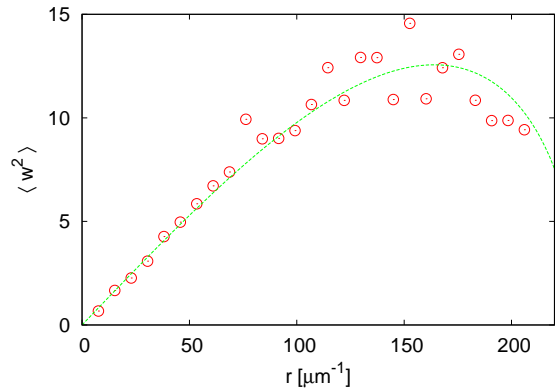}
\caption{
Variance of the winding number $\langle w^2(r) \rangle$ along the
circumference of a circle of radius $r$ for sudden quench, where the
parameters are the same as those in Fig.~\ref{f:sudden}.
The data for each $r$ is taken when $G_{\rm T}(r)$ exceeds 0.1.
The average $\langle \cdots \rangle$ is taken over 400 runs of simulations
for the different initial states produced by random numbers.
The dashed curve is the least square fit of Eq.~(\ref{wrdep}).
}
\label{f:rdep}
\end{figure}
We consider the $r$-dependence of the spin winding number $w(r)$.
According to the KZ theory, the number of domains along the circle of
radius $r$ is $\sim r / \xi_{\rm corr}$ and hence $w^2(r) \sim r /
\xi_{\rm corr}$.
Substituting $q = 0$ and $q_{\rm c} = 2 |c_1| n_{\rm TF}(r)$ into
$\xi_{\rm corr}$ in Eq.~(\ref{xcorr2}), where $n_{\rm TF}(r) \propto
R_{\rm TF}^2 - r^2$ with $R_{\rm TF}$ being the Thomas--Fermi radius, we
obtain
\begin{equation} \label{wrdep}
w^2(r) \propto r \sqrt{R_{\rm TF}^2 - r^2}.
\end{equation}
To compare Eq.~(\ref{wrdep}) with the numerical simulation, we perform
many runs of time evolution with different initial random noises, and take
the average of $w^2(r)$ with respect to the runs, which is shown in
Fig.~\ref{f:rdep}.
Since the time at which the magnetization emerges depends on $r$, each
$w(r)$ is calculated when $G_{\rm T}(r)$ in Eq.~(\ref{gtr}) exceeds a
certain value (0.1 in Fig.~\ref{f:rdep}).
The numerical result and Eq.~(\ref{wrdep}) (circles and dashed curve in
Fig.~\ref{f:rdep}, respectively) are in good agreement, where the fitting
parameter is only the proportionality coefficient in Eq.~(\ref{wrdep}).

\subsection{Gradual quench}
\label{s:slow}

We next consider the case of gradual quench of the magnetic field.
The quadratic Zeeman energy $q$ is linearly decreased in the time scale
$\tau_{\rm Q}$ as
\begin{equation} \label{qt}
q(t) = q_0 (1 - t / \tau_{\rm Q}),
\end{equation}
for $0 < t < \tau_{\rm Q}$ and $q(t) = 0$ for $t \geq \tau_{\rm Q}$.
As seen in the previous subsection, the critical value $q_{\rm c}(r)$ for
magnetization depends on the position $r$, and $q_0$ is chosen to be the
maximum of $q_{\rm c}(r)$.
We define the time $T(r)$ at which $q$ reaches a local critical value as
$q(T(r)) = q_{\rm c}(r)$.
The magnetization at position $r$ is expected to emerge at $t = T(r) +
\Delta t(r)$ satisfying~\cite{Zurek}
\begin{equation} \label{tsimtau}
\Delta t(r) \sim \tau(r, t).
\end{equation}
Using Eqs.~(\ref{tau1}) and (\ref{tsimtau}) with $q_0 \Delta t(r) /
[q_{\rm c}(r) \tau_{\rm Q}] \ll 1$, we obtain
\begin{equation}
\Delta t(r) \sim \left[ \frac{\hbar^2}{q_{\rm c}(r) q_0} \right]^{1/3}
\tau_{\rm Q}^{1/3}.
\end{equation}
Substituting this time into Eq.~(\ref{xi1}), we obtain the
$\tau_{\rm Q}$-dependence of the correlation length as
\begin{equation}
\xi_{\rm corr}(r) \sim \sqrt{\frac{\hbar^2}{M q_0}}
\left[ \frac{q_{\rm c}(r) q_0}{\hbar^2} \right]^{1/6} \tau_{\rm Q}^{1/3}.
\end{equation}
The winding number thus obeys
\begin{equation} \label{wpow}
w^2(r) \propto \tau_{\rm Q}^{-1/3}.
\end{equation}

\begin{figure}[tb]
\includegraphics[width=8cm]{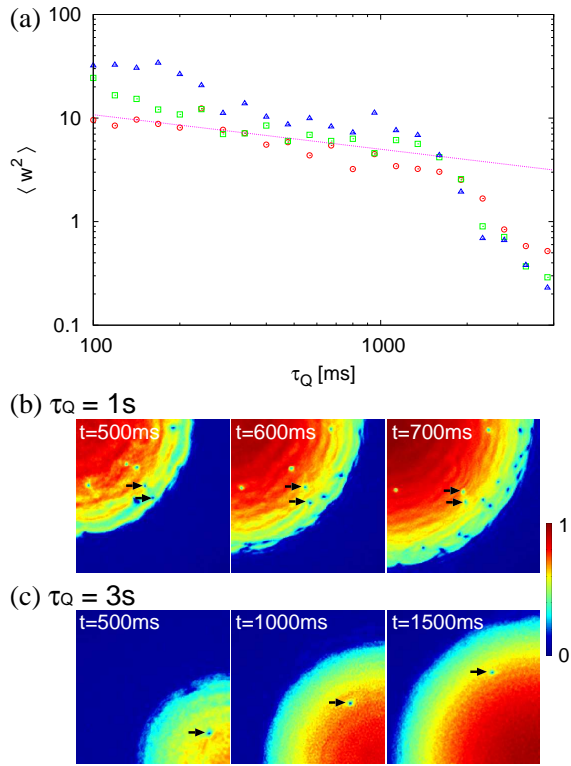}
\caption{
(a) Variance of the winding number $\langle w^2(r) \rangle$ along the
circumference of a circle of radius $r = 200$ $\mu{\rm m}$ (circles), 
$r = 250$ $\mu{\rm m}$ (squares), and $r = 300$ $\mu{\rm m}$ (triangles)
for gradual quench given by Eq.~(\ref{qt}).
The data for each $r$ is taken when $G_{\rm T}(r_{\rm w})$ exceeds 0.1.
The average $\langle \cdots \rangle$ is taken over 100 runs of simulations
for the different initial states produced by random numbers.
The dashed line is proportional to $\tau_{\rm Q}^{-1/3}$.
The number of atoms is $N = 10^8$.
(b), (c) Snapshots of the transverse magnetization $|F_+(\bm{r}, t)|$ for
$\tau_{\rm Q} =$ 1 s and 3 s.
The arrows trace the vortex motion.
The unit of $|F_+(\bm{r}, t)|$ is $3.4 \times 10^{14}$ ${\rm m}^{-2}$.
The field of view of each panel is $300 \times 300$ $\mu {\rm m}$.
See the supplementary data files for the movies showing the dynamics.
}
\label{f:slow}
\end{figure}
Figure~\ref{f:slow} (a) shows $\langle w^2(r) \rangle$ obtained by
numerical simulations of the GP equation (\ref{GP}).
The variance of the winding number $\langle w^2(r) \rangle$ is roughly
proportional to $\tau_{\rm Q}^{-1/3}$ for $100$ ${\rm ms} \lesssim
\tau_{\rm Q} \lesssim 1$ s, which agrees with the above theoretical
argument (\ref{wpow}).
For $\tau_{\rm Q} \lesssim 100$ ms, the assumption of $t / \tau_{\rm Q}
\ll 1$ is violated;
$\langle w^2(r) \rangle$ approaches the values for sudden quench in the
limit of $\tau_{\rm Q} \rightarrow 0$.
For $\tau_{\rm Q} \gtrsim 1$ s, $\langle w^2(r) \rangle$ significantly
deviates from $\tau_{\rm Q}^{-1/3}$ and steeply drops in Fig.~\ref{f:slow}
(a).
To understand this behavior, we compare the dynamics of $|F_+(\bm{r})|$
for $\tau_{\rm Q} = 1$ s and $\tau_{\rm Q} = 3$ s shown in
Figs.~\ref{f:slow}(b) and \ref{f:slow}(c).
When $\tau_{\rm Q} = 1$ s, new spin vortices are produced one after
another as the magnetization grows outward.
When $\tau_{\rm Q} = 3$ s, by contrast, the spin vortex created around the
center is pushed outward, and no new spin vortices are created at the
front of the magnetization growth.

In the dynamics in Figs.~\ref{f:slow}(b) and \ref{f:slow}(c), there are
two characteristic velocities: the velocity $v_{\rm m}$ at which the
magnetization front spreads out and the sound velocity $v_{\rm s}$ of the
spin wave.
The former is roughly obtained from
\begin{equation}
q(t) = q_{\rm c}(r_{\rm M}(t)),
\end{equation}
where $r_{\rm M}(t)$ is the radius of the magnetization front.
Using the Thomas--Fermi density distribution, the right-hand side is
$q_{\rm c}(r_{\rm M}(t)) = 2|c_1| n_{\rm TF}(r_{\rm M}(t)) \simeq q_0 [1 -
r_{\rm M}^2(t) / R_{\rm TF}^2]$.
The velocity $v_{\rm m}$ is thus given by
\begin{equation} \label{vm}
v_{\rm m} = \frac{dr_{\rm M}(t)}{dt} =
\frac{R_{\rm TF}}{2 \sqrt{\tau_{\rm Q} t}}. 
\end{equation}
The transverse spin wave for the broken axisymmetry state (\ref{ba}) is a
phonon-like mode in the limit of $k \rightarrow 0$, whose velocity is
given by~\cite{Kawaguchi,Murata} 
\begin{equation} \label{vs}
v_{\rm s} = \sqrt{\frac{q}{2M}},
\end{equation}
where $q(t)$ in Eq.~(\ref{qt}) should be used on the right-hand side.
If $v_{\rm m}$ is always faster than $v_{\rm s}$, the region that is going
to magnetize is causally disconnected with the magnetized region, and
therefore the KZ mechanism works.
It follows from Eqs.~(\ref{vm}) and (\ref{vs}) that this condition is
satisfied for
\begin{equation}
\tau_{\rm Q} < \sqrt{\frac{2M}{q_0}} R_{\rm TF}.
\end{equation}
For the parameters in Fig.~\ref{f:slow}, the right-hand side of this
inequality is $\simeq 1.7$ s, which agrees well with the time at which the
plots in Fig.~\ref{f:slow}(a) deviates from $\tau_{\rm Q}^{-1/3}$.

\subsection{Gradual quench with a plug potential}
\label{s:plug}

\begin{figure}[tb]
\includegraphics[width=8cm]{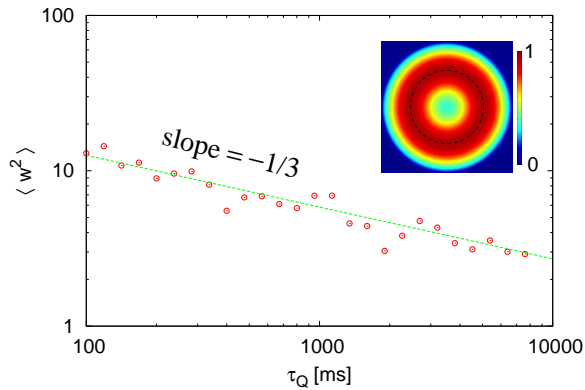}
\caption{
Variance of the winding number $\langle w^2(r) \rangle$ along the
circumference of a circle of radius $r = 250$ $\mu{\rm m}$ (dashed circle)
for gradual quench given by Eq.~(\ref{qt}).
The potential has a form of Eq.~(\ref{plug}).
The data for each $r$ is taken when $G_{\rm T}(r_{\rm w})$ exceeds 0.1.
The average $\langle \cdots \rangle$ is taken over 100 runs of simulations
for the different initial states produced by random numbers.
The dashed line is least square fit of the plots by using a function
proportional to $\tau_{\rm Q}^{-1/3}$.
The number of atoms is $N = 10^8$.
The inset shows the initial density profile with the unit of density
$2.4 \times 10^{14}$ ${\rm m}^{-2}$ and the field of view
$900 \times 900$ $\mu {\rm m}$.
}
\label{f:plug}
\end{figure}
In the previous subsection, we showed that the KZ scenario breaks down
when the magnetized region expands slowly.
This is because the region that is going to magnetize is causally
connected to the initially magnetized region at the trap center.
To eliminate this effect, we remove atoms around the trap center by adding
a plug potential as
\begin{equation} \label{plug}
V_{\rm trap}(\bm{r}) = \frac{1}{2} M \omega^2 r^2 + A e^{-r^2 / d^2},
\end{equation}
where the values of the parameters are chosen to be $A = 1500 \hbar
\omega$ and $d = 222$ $\mu{\rm m}$.
For these parameters, the potential $V_{\rm trap}(\bm{r})$ has a minimum
at $r \simeq 250$ $\mu{\rm m}$ and the atomic density becomes maximal
around this radius.
As a result, the magnetization starts from the annulus around $r \simeq
250$ $\mu{\rm m}$.
Therefore, magnetic domains on this radius are always causally
disconnected, and the KZ mechanism is expected to work even for large
$\tau_{\rm Q}$.
Figure~\ref{f:plug} shows the results of the numerical simulations.
The density at $r \simeq 250$ $\mu{\rm m}$ is almost the same as the
density at the same radius in the system of Fig.~\ref{f:slow}, and
$\langle w^2(r) \rangle$ is similar to the corresponding data in
Fig.~\ref{f:slow}(a) (squares) for $\tau_{\rm Q} \lesssim 1$ s.
However, $\langle w^2(r) \rangle$ in Fig.~\ref{f:plug} obeys the KZ
scaling $\tau_{\rm Q}^{-1/3}$ even for $\tau_{\rm Q} \gtrsim 1$ s, as
expected.

\section{Conclusions}
\label{s:conc}

We have investigated the spin vortex formation due to the KZ mechanism in
a quenched ferromagnetic BEC confined in a trapping potential.
Since the atomic density is inhomogeneous in a harmonic trap, the spin
correlation length depends on the radius $r$.
In fact, the numerical simulations showed that the spin winding number
depends on $r$, which was in good agreement with the theoretical prediction
(Fig.~\ref{f:rdep}).
When the quadratic Zeeman energy $q$ is gradually quenched with the time
scale $\tau_{\rm Q}$, the magnetized region gradually expands from the
center to the periphery of the atomic cloud.
If the expansion velocity is much faster than the spin wave velocity, the
system exhibits the KZ scaling law, and if the former is slower than the
latter, the KZ scenario breaks down (Fig.~\ref{f:slow}).
When a plug potential is added to the harmonic trap, the geometry of the
system is changed and the KZ power law can be observed over a wide range
of $\tau_{\rm Q}$ (Fig.~\ref{f:plug}).

\begin{acknowledgements}
This work was supported by Grants-in-Aid for Scientific
Research (No.\ 22103005, No.\ 22340114, No.\ 22340116, No.\ 22740265, and
No.\ 23540464) from the Ministry of Education, Culture, Sports, Science
and Technology of Japan.
YK acknowledges the financial support from Inoue Foundation.
\end{acknowledgements}

\end{document}